\documentclass[aps,twocolumn,showpacs]{revtex4-1}
\usepackage{graphics}
\usepackage{graphicx}
\usepackage{latexsym}
\usepackage{amsmath}
\usepackage{dsfont}
\def\beqn{\begin{eqnarray}}
\def\eeqn{\end{eqnarray}}
\def\nnu{\nonumber \\}

\def\uni{ {\mathbf{i}}}
\def\hw{\hbar \omega}
\def\hw4{ \frac {\hbar \omega}{4}}

\def\bp{b^\dagger}

\begin{document}
\title{Optimal Spin Squeezed Steady State induced by the dynamics of non-hermtian hamiltonians.}
\vskip2cm
\author{Ram\'\i rez R. $^{a)}$}
\author{Reboiro M. $^{b)}$ \footnote{e-mail: reboiro@fisica.unlp.edu.ar}}
\affiliation{{\small\it $^{a)}$Department of Mathematics, University of La Plata}
{\small \it La Plata,Argentina}}
\affiliation{{\small\it $^{b)}$IFLP, CONICET-Department of Physics, University of La Plata}
{\small \it La Plata, Argentina}}
\date{\today}

\begin{abstract}
In this work, we study the time evolution of a coherent spin state under the action of a non-hermitian hamiltonian. The hamiltonian is modeled by a one-axis twisting term plus a Lipkin-type interaction. We show that when the Lipkin interaction is switched on, depending on the relative values of the coupling constants, the initial state evolves into a steady squeezed state which minimizes the Uncertainty Relations, Intelligent Spin State. We apply this result to look for the generation of an steady intelligent spin state from an ensemble of nitrogen vacancy colour centers in diamond coupled to a mechanical resonator.
\end{abstract}


\maketitle

key words: non-hermitian dynamics, optimal spin squeezing, one-axes twisting and Lipkin-type interactions.

\section{Introduction}

The one-axis-twisting (OAT) and the two-axis-twisting (TAT) mechanisms have been introduced by Kitagawa and Ueda \cite{kitagawa} to establish the concept of spin squeezing states and the fundamentals for their generation. From the theoretical point of view, squeezing is closely related to the analysis of Heisenberg Uncertainty Relations. It means that given a physical system, one may be interested in the minimization of the fluctuation of an observable at the expense of the increment of the fluctuation of the conjugate variable.

Since the pioneering  work of Kitagawa and Ueda \cite{kitagawa}, many authors have contributed to the understanding \cite{nori,nemoto,sorensen} and to the experimental achievement of spin squeezing in atomic systems \cite{bec-exp-1,bec-exp-2,bec-exp-3}. Recently, the interest in the study of these mechanisms has been renewed \cite{bec-0,bec-1,bec-2,tat-0,tat-2,tat-4,tat-1,tat-3,oat-1,oat-5,lipkin1,oat-4,oat-2,oat-3}. The characterization of spin squeezing is relevant in the analysis of potential candidates to be used in the architecture of quantum computing devices \cite{arxiv}. In a series of works, it has been reported the generation of steady squeezed states in dissipative spin systems \cite{oat-3,torre,disi-0,disibis,disi-1,disi-4,disi-5,disi-3,disi-2,disi-6,nosaphys}. As an example, we can mention the analysis of phase coherence and spin squeezing of collective spin in systems governed by OAT Hamiltonian with decay \cite{tat-1,oat-1,oat-5,oat-4,oat-2,oat-3,sorensen} or in systems governed by non-Hermitian Lipkin-Meshkov-Glick hamiltonian (LMG) \cite{tat-0,disi-0,disibis}. Similar results were found in the study of the behavior of dissipative  hybrid systems \cite{disi-3,disi-1,nosaphys,zhu,zhubis,marco,nv-qb-1,qb-nv,photons-coupling,ma-3}. The reported works can be taken as an indication that non-hermitian dynamics can be used to improve the achievement of squeezing in different spin system.

Moreover, the search for spin squeezed states with minimum uncertainty relations has given rise to the notion of Intelligent Spin State (ISS) \cite{iss1}.
The first references in the literature to intelligent states there is the paper of C. Argone and co-workers \cite{iss1}. A considerable amount of work was devoted to the study of both the properties of intelligent spin states \cite{iss2} as well as to the construction of such states \cite{iss3,iss4,iss5,iss6,nosiss}. In this work, we analyse the generation of a steady ISS in a system of spins interacting through a non-Hermitian OAT Hamiltonian plus a LMG interaction.
As a physical application, we propose to search for steady ISS in diamond nanostructures \cite{ma-1,ma-2,ma-20,mab,phonon1,coupling00}.

Among other proposals, nitrogen-vacancy (NV) centers in diamond may be useful in solid quantum information processing due to their long coherence time and to the high feasibility in their manipulation \cite{zhu,zhubis,marco,nv-int-2,nv-1,nv-int-1,nv-int-new,hybrid-10,hybrid-11}.
The generation of entanglement among NV centers in diamond has been achieved by different mechanisms. The coupling of pairs of NV centers have been obtained directly by dipole-dipole interaction \cite{wrachtrup1,wrachtrup2}. The coherent coupling of an ensemble of NV centers to a superconducting resonator have been reported in \cite{kubo}. Also, the coupling of two separated NV electron spin ensembles in a cavity quantum electrodynamics system has been observed recently \cite{prl118}. Another novel mechanism to generate long-range spin-spin interactions in NV centers in diamond has been proposed in \cite{phonon1}. In this scheme the interaction among NV centers is
mediated by their coupling via strain to the vibrational mode of a diamond mechanical nanoresonator. The Authors of \cite{phonon1} have probed that this phonon-mediated effective spin-spin interactions can be used
to generate squeezed states of the spin ensemble. In the same direction, the Authors of \cite{ma-1,ma-2} have shown that under the action of an effective phonon-induced spin-spin interaction for the ensemble of NV color centers in diamond, the initial state evolves into a steady state that behaves as a squeezed state. In this work, we model the interaction of an ensemble of NV centers in diamond coupled to a mechanical resonator by an effective OAT plus LGM effective hamiltonian for the NV centers. We investigate the possibility of the generation of an steady ISS from the time evolution of an initial prepared coherent state under the action of this effective hamiltonian.

The work is organized as follows. The details of the general formalism are presented in Section \ref{formalism}.
The results of the calculations are presented and discussed in Section \ref{results}.
In Section \ref{numbers}, we present the numerical  results that we have obtained from the exact diagonalization of the proposed Hamiltonian. In Sections \ref{otwist} and \ref{su11} we study some analytical results, so to better understand the mechanism of generation of a steady ISS. In Section \ref{otwist}, the time evolution and the asymptotic behavior of an initial coherent state under the action of a non-hermitian OAT Hamiltonian is discussed. In Section \ref{su11}, we study the behaviour of the system, when the LMG interaction is taken into account, by performing a boson mapping and keeping terms to dominant order in the number of spins. In doing so, we explore the behavior of the steady state on the different parameters of the model. In section \ref{application} we propose a scheme to couple an ensemble of NV centers to a mechanical resonator, so that the system can be model by an effective phonon-mediated interaction, which consists of a OAT plus a LGM interaction. We discuss the generation of a steady ISS for this effective model. Our conclusions are drawn in Section \ref{conclusions}.

\section{Formalism}\label{formalism}

Let us consider a general collective system consisting of $2 S$ elementary $1/2$-pseudo-spins \cite{zhu,zhubis,marco,qb-nv}.
The collective pseudo-spin of the system, ${\bf S}= \left( ~S_x,~S_y,~S_z \right)$, is governed by the cyclic commutation relations $\left[~S_i,~S_j\right]=~{\rm \bf i}~\epsilon_{ijk}~S_k$, where the suffixes $i,j,k$ stand for the components of the spin in three orthogonal directions and $\epsilon_{ijk}$ is the Levi-Civita symbol. We shall assume that the physical properties of the system can be modeled by a Hamiltonian of the form

\beqn
H & = & H_{OAT}+H_{LMG}+H_{\gamma}, \nonumber \\
H_{OAT}&=& \chi \, S_z^2, \nonumber \\
H_{LMG} & = & = V (S_x^2-S_y^2),\nonumber \\
H_{\gamma} & = & \left( \epsilon-{\rm \bf i} \gamma \right) \left(S_z+ S \right).
\label{hota}
\eeqn
The term $H_{OAT}$ of the Hamiltonian of Eq. (\ref{hota}) is a one-axis twisting mechanism with coupling constant $\chi$, while term $H_{LMG}$ stands for a Lipkin-type interaction \cite{lipkin,newlmg1}. In addition, we shall assume that the particles of the system have a finite lifetime, which is given by the line-width $\gamma$. This effect can be model by the non-hermitian term $H_{\gamma}$ \cite{disi-0}.

From the theoretical point of view, different physical systems can be modeled by hamiltonians closely related to one proposed in Eq.( \ref{hota} ), i.e.  a system of two-component atomic condensates \cite{oat-1,tat-4,bec-1,bec-2}, or an ensemble of NV centers coupling via a mechanical resonator \cite{ma-1,ma-2,ma-20,phonon1,coupling00}.

The Hamiltonian of Eq. (\ref{hota}) can be diagonalized exactly in the basis of states ${\mathcal A}_k=\{ |k \rangle \}$, with

\beqn
|k \rangle=| S,~-S+ k  \rangle = \left[ \frac {(2~S-k)!}{(2~S)! k!}\right]^{1/2} S_+^k ~ |S, ~-S \rangle.
\label{base}
\eeqn
In this basis

\beqn
{\bf S}^2 |k \rangle =S(S+1)~|k \rangle,~~~S_z| k \rangle=(-S+k) | k \rangle.
\eeqn

\subsection{Time Evolution.}\label{tevol}

In writing the Hamiltonian of Eq. (\ref{hota}), we have followed the projection operator formalism of Feshbach \cite{feschbach} to introduce the non-hermtian dynamics of the system.

As the Hamiltonian of Eq. (\ref{hota}) is non-hermitian, we have

\beqn
H |\widetilde{\phi}_\alpha \rangle= E_\alpha |\widetilde{\phi}_\alpha \rangle,
\eeqn
and
\beqn
H^\dagger | \overline{\psi}_\alpha \rangle = {\overline{E}}_\alpha | \overline{\psi}_\alpha \rangle.
\eeqn
Both sets of eigenstates, ${ \mathcal A }_H= \{ | {\widetilde \phi}_\alpha \rangle \}$ and ${\mathcal A}_{H^\dagger}= \{ |{\overline \psi}_\alpha \rangle \}$, are non-orthonormal basis of the Hilbert space, ${\mathcal H}$. It is straightforward to prove \cite{faisal,rotter1,rotter2} that
\beqn
{\overline{E}}_\alpha={\widetilde{E}}^*_\alpha,
\eeqn
and that the set $\{ |\overline{\psi}_\alpha \rangle, |\widetilde{\phi}_\beta \rangle \}$ forms a bi-orthonormal basis of ${\mathcal{H}}$, with

\begin{equation}
\langle {\overline \psi}_\alpha | {\widetilde \phi}_\beta \rangle = \delta_{\alpha \beta}.
\label{deltaBiON}
\end{equation}

Clearly, the spectrum of the Hamiltonian of Eq. (\ref{hota}) depends on the value of the coupling constants \cite{arxiv1}. If $\epsilon=0$, the hamiltonian $H$ of Eq. (\ref{hota}) is a quasi-hermitian operator, and its spectrum has complex pair conjugate eigenvalues. It means that $H$ is iso-spectral to $H^\dagger$.  Otherwise, the spectrum of $H$ contains complex (non-pair-conjugate) eigenvalues, and the eigenvalues of $H^\dagger$ are complex conjugate to the eigenvalues of $H$.

In the basis ${\mathcal A}_k$, a general initial state can be written as

\beqn
| I \rangle= \sum_k ~ c_k ~ | k \rangle.
\label{ini00}
\eeqn
In terms of the basis formed by the eigenvectors of $H$ the initial state is given by
\beqn
| I \rangle & = & \sum_\alpha ~ \widetilde{c}_\alpha ~ |\widetilde{\phi}_ \alpha \rangle, \nonumber \\
\widetilde{c}_\alpha & = & \sum_k~ (\Upsilon^{-1})_{\alpha k}~ c_k,
\label{ini0}
\eeqn
with $\Upsilon$ the transformation matrix from basis ${\mathcal A}_k$ to basis ${\mathcal{A}}_H $.
We shall assume that the initial state is normalized,
that is $\langle {I} | {I} \rangle=1$. The initial state of Eq.(\ref{ini0}) evolves in time as

\beqn
| I(t) \rangle & = & {\rm e}^{- i H t} | I \rangle, \nonumber \\
  & = & \sum_\alpha ~ {{\widetilde c}_\alpha(t)} ~ |\widetilde{\phi}_\alpha \rangle.
\label{init}
\eeqn
If $H$ can be diagonalized, $\widetilde{c}_\alpha(t)$ is given by $\widetilde{c}_\alpha(t)={\rm e}^{- i \widetilde{E}_\alpha t}~\widetilde{c}_\alpha$.

In order to work with the basis formed by the eigenstates of $H$,  $\mathcal{A}_{H}$, to calculate the expectation value of a given observable, $\widehat{o}$, we have to equipped the linear vector space with an scalar product. The reader is kindly refer to \cite{arxiv1} and references therein. That is, we look for a metric operator $\mathcal{S}$, i.e. an operator which is auto-adjoint and positive definite. The Hilbert space ${\mathcal H}$ equipped with the scalar product $\langle  {\bf f} | {\bf g} \rangle_{\mathcal S}=\langle  {\bf f} | {\mathcal S}{\bf g} \rangle_{\mathcal S}$ is the new physical linear space ${\mathcal H}_{\mathcal S}=(\mathcal{H}, \langle .| . \rangle_{\mathcal S})$.
In terms of the eigenvectors of the symmetry operator $\mathcal{S}$, the initial state reads

\beqn
|I(t) \rangle & = &
\sum_\beta~
{\overset{\approx}{c}}_\beta(t)~|{\overset{\approx}{\phi}}_\beta \rangle,\nonumber \\
{\overset{\approx}{c}}_\beta(t)& = & \sum_{\alpha}~
(\Upsilon'^{-1})_{\beta \alpha}~  \widetilde{c}_\alpha(t),
\label{inits}
\eeqn
with $\Upsilon'$ being the transformation matrix from  the  basis  ${\mathcal A}_H$ to the basis ${\mathcal A}_S$.
We are know in condition of evaluate
the mean value of an operator $\widehat{o}$ as a function of time as
\beqn
\langle \widehat{o}(t) \rangle & = & {\langle I (t)| \widehat{o}|I(t) \rangle}_{\mathcal{S}} \nonumber \\
& = & \sum_{\alpha \beta}~ {\overset{\approx}{c}}_\alpha(t) {\overset{\approx}{c}}^*_{\beta}(t)~
 \langle {\overset{\approx}{\phi}}_\beta  \mid \widehat{o} \mid {\overset{\approx}{\phi}}_\alpha \rangle_{\mathcal{S}}.
\eeqn

As reported in \cite{arxiv1}, the form of the metric operator depends on the spectrum of $H$. It can be summarized as follows.

If the spectrum of $H$ contains complex pair conjugate eigenvalues, there exists a symmetry self-adjoint operator such that ${\mathcal S_K} H = H^\dagger {\mathcal S_K}$. It reads

\beqn
{\mathcal S_K}
& = & \sum_{j \le i}^{N_{max}}
~ \delta( \overline{E}_j-\bar{E}^*_i) ~ \left( \alpha_{j}|  \bar{\psi}_{j} \rangle  \langle \bar{\psi}_{i}|+
~ \alpha^*_{j}|  \bar{\psi}_{i} \rangle  \langle \bar{\psi}_{j}| \right ). \nonumber \\
\eeqn
This operator is not positive define, so that we make use of the formalism of Krein Spaces. After the diagonalization of ${\mathcal S}$, we have ${\mathcal S_K}=R D R^{-1}=R D_+ R^{-1}+ R D_- R^{-1}=S_{K+}+S_{K-}$, with $D_+$ the diagonal matrix with positive elements and $D_-$ the diagonal matrix with negative entries.
Finilly, the metric operator is given by ${\mathcal S}=S_{K+}-S_{K-}$.

If the non-hermitian Hamiltonian $H$ has real eigenvalues or some eigenvalues are complex (non-pair-conjugate), the metric operator is given by

\beqn
{\mathcal S} = \sum_{j=1}^{N_{max}} ~ |  \overline {\psi}_{j} \rangle  \langle \overline{\psi}_{j}|.
\label{opSAg}
\eeqn

\subsection{Spin-Squeezing Parameter and Intelligent Spin States.}\label{squeezing}
Spin-squeezed-states are quantum-correlated states  with
reduced fluctuations in one of the components of the total spin. Following the work of Ueda and Kitagawa \cite{kitagawa}, we shall
define a set of orthogonal axes $\{ {\bf n_{x'}}, {\bf n_{y'}},
{\bf n_{z'}} \}$, such that ${\bf n_{z'}}$ is the unitary vector pointing along the direction
of the total spin $<{\bf S }>$ .  We shall fix the direction ${\bf n_{x'}}$ by looking at
the minimum value of $\Delta^2S_{x'}$. The Heisenberg  Uncertainty Relation reads

\beqn
\Delta^2 S_{y'} ~\Delta^2 S_{x'}~& \ge & ~\frac 14 |<{\bf S}>|^2.
\label{hur}
\eeqn
We define the squeezing parameters \cite{kitagawa} as

\begin{eqnarray}
\zeta^2_{x'}  =  \frac {2 \Delta^2 S_{x'} }{|<\bf{S} >|},~
\zeta^2_{y'}  =  \frac {2 \Delta^2 S_{y'} }{|<\bf{S} >|}.
\label{sqx}
\end{eqnarray}
The state is squeezed in the $x'$-direction if $\zeta^2_{x'}<1$ and $\zeta^2_{y'}>1$. So defined, the
parameters of Eq. (\ref{sqx}) are su(2) invariant
\cite{luis}.

When the minimum value of the Heisenberg Uncertainty Relation, Eq. (\ref{hur}), is achieved and $\zeta^2_{x'}<1$, the state is
called Intelligent Spin State \cite{iss1,iss2,iss3,iss4,iss5,iss6}.

\section{Results and discussions} \label{results}

Let us first present and discuss general results obtained for the time evolution of a Coherent Spin State (CSS) \cite{hecht} through the action of the Hamiltonian of Eq.(\ref{hota}). The initial state has the form

\begin{eqnarray}
|I (\theta_0,\phi_0)\rangle= {\cal N} \sum_{k=0}^{2 S}~ z(\theta_0,\phi_0)^k \left (\begin{array}{c} 2 S \\ k \end{array} \right)^{1/2} |k\rangle , \label{istate}
\end{eqnarray}
with $ z(\theta_0,\phi_0) ={\rm e}^{-i \phi_0} \tan(\theta_0/2)$. The angles $(\theta_0,\phi_0)$ define the direction
$\vec{n}_{0}=(\sin{\theta_0} \cos{\phi_0},\sin{\theta_0}
\sin{\phi_0},\cos{\theta_0})$, such that $\vec{S} \cdot \vec{n}_0 |I\rangle=-S | I\rangle$ \cite{hecht}.

 We shall begin with the analysis, in Section \ref{numbers}, of the numerical results obtained from the exact diagonalization of the Hamiltonian of Eq.(\ref{hota}). We shall complement these results with the analytical ones of Sections \ref{otwist} and \ref{su11}. Finally, we shall investigate the possibility of generating a steady Intelligent Spin States in diamond nanostructures, \ref{application}.

\begin{figure}[h!]
\includegraphics[width=6cm]{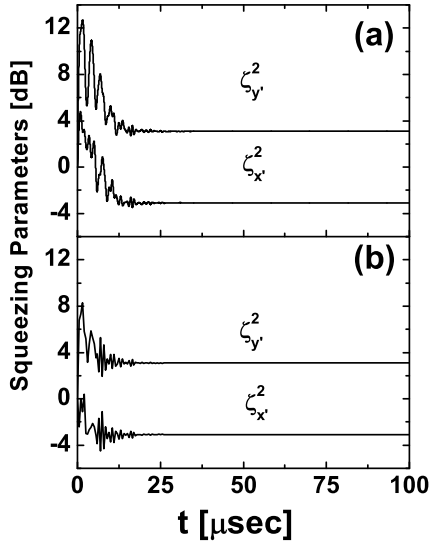}
\caption{Squeezing parameters, $\zeta^2_{x'}$ and $\zeta^2_{y'}$, as a function of time, for the system model by the Hamiltonian of Eq.(\ref{hota}), in units of [dB]. The system consists of N=45 spins. The parameters of the model have been fixed to the values $\eta=0.6$, $\gamma=2 \times 10^{-5}$ [GHz].
In Insets (a) and (b) are displayed the results obtained when the initial coherent state is prepared with $\left( \theta_0, \phi_0 \right)=(\pi/4,0)$ and $(\pi/8,0)$, respectively. } \label{fig:fig1}
\end{figure}

\begin{figure}[h!]
\includegraphics[width=6cm]{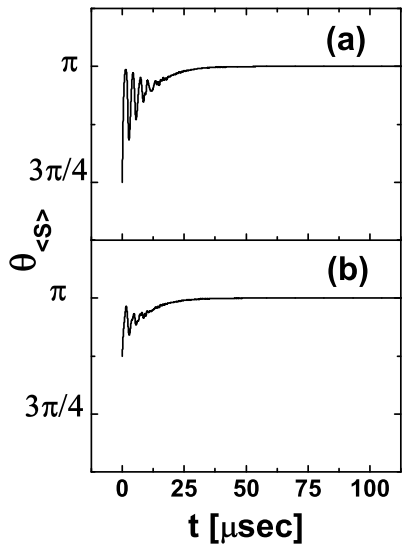}
\caption{Polar angle of the unit vector along the direction of the mean value of the total spin, $\theta_{<{\mathbf S}>}$, as a function of time. The parameters are the same as in Figure 1.
In Insets (a) and (b) are displayed the results obtained when the initial coherent sate is prepared with $\left( \theta_0, \phi_0 \right)=(\pi/4,0)$ and $(\pi/8,0)$, respectively. } \label{fig:fig2}
\end{figure}

\begin{figure}[h!]
\includegraphics[width=6cm]{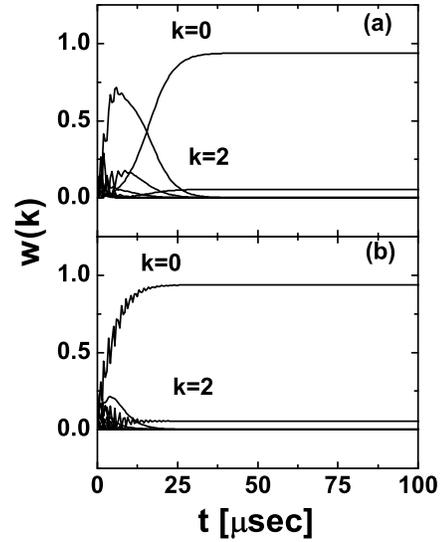}
\caption{Contribution of the $k-$th state of the basis ${\mathcal A}_k$ to the state $|I(t) \rangle$ of Eq.(\ref{ini00}), as a function of time. The parameters are those of Figures 1 and 2.
In Insets (a) and (b) are displayed the results obtained when the initial coherent state is prepared with $\left( \theta_0, \phi_0 \right)=(\pi/4,0)$ and $(\pi/8,0)$, respectively. } \label{fig:fig3}
\end{figure}

\begin{figure}[h!]
\includegraphics[width=8cm]{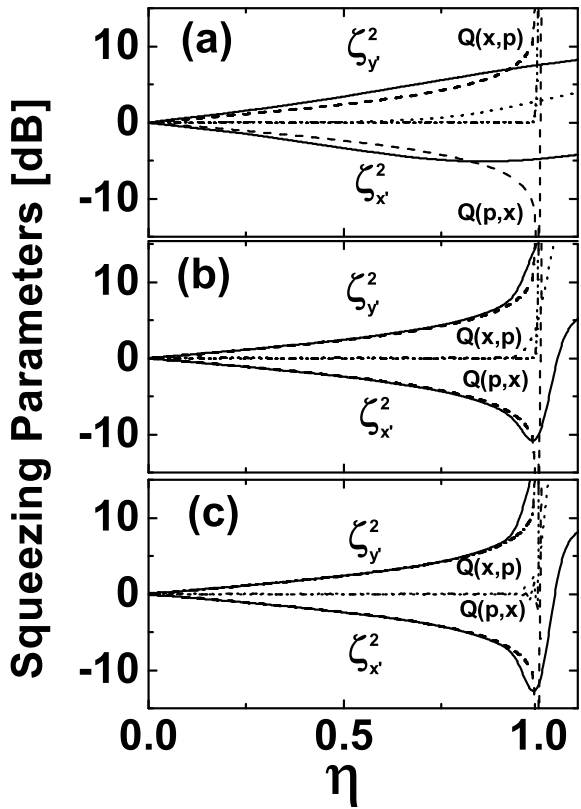}
\caption{Dependence, as a function of the relative coupling constant $\eta$, of the Squeezing Parameters of the steady state ($t>> T_c$, $t=120$ [$\mu$ sec]), in units of [dB].
In Insets (a), (b) and (c) we plot the results obtained for ensembles with $N=5$, $N=45$ and $N=101$ spins, respectively. We have fixed $\gamma=2 \times 10^{-5}$ [GHz].
Solid lines are used to shown the results which we have obtained for the squeezing parameters from the exact diagonalization of the Hamiltonian of Eq.(\ref{hota}), $\zeta^2_{x'}$ and $\zeta^2_{y'}$ of Eq.(\ref{sqx}), for an initial coherent state  with $(\theta_0,~\phi_0)=(\pi/4,~0)$, Eq. (\ref{istate}). Dashed-lines correspond to the results which we have obtained by applying the boson approximation of section \ref{su11}, $Q(x,p)$ and $Q(p,x)$ of Eq.(\ref{sqbos}). In this case, the initial state of Eq. (\ref{inibos}) consists of $5$ particles in mean value for Inset (a), and of $45$ and $101$ particles in mean value for Inset (b) and (c), respectively. With dotted-line and with dashed-dotted-line we present the results for the product of the squeezing parameters in the exact and in the approximate case, respectively.  } \label{fig:fig4}
\end{figure}

\subsection{Exact Numerical Results.}\label{numbers}

As it has been said before, we shall present results corresponding to the time evolution of the initial coherent state of Eq.(\ref{istate}) under the action of the Hamiltonian of Eq. (\ref{hota}), by performing the exact diagonalization of the hamiltonian in the basis Eq. (\ref{base}). In doing so, we shall describe the behavior of the system in terms of the relative coupling constants

\beqn
\eta   & = & \frac{ |2 S V| }{ |\epsilon -2 S \chi| }, \nonumber \\
\Gamma & = & \frac{ \gamma }{ |\epsilon -2 S \chi| },
\label{ceta}
\eeqn
and of the parameter
\beqn
\Xi^2 &= & \eta^2+\Gamma^2.
\label{xi}
\eeqn

In Figures 1, 2 and 3 we present numerical results for a system consisting of $45$ spins. We assume that the system has a characteristic time of coherence of the order of $T_C \approx 100$ [$\mu$ sec ] \cite{zhu,nv-ct1,nv-ct2,nv-ct3}, which is consisting with a value for the line-width of the states
of $\gamma=2 \times 10^{-5}$ [GHz], relative to the coupling constant $\chi$.

In Figure 1, we show the behaviour of the squeezing parameters of Eq. (\ref{sqx}), $\zeta^2_{x'}$ and $\zeta^2_{y'}$, as a function of time. We have fixed the coupling relative constant $\eta$ to the value $\eta=0.6$.
In Insets (a) and (b) we have displayed the results obtained when the initial coherent state is prepared with $\left( \theta_0, \phi_0 \right)=(\pi/4,0)$ and with $(\pi/8,0)$, respectively. At intermediate times, the pattern of squeezing depends on the value of $\theta_0$. Initial states with $\theta_0$ smaller than $\pi/4$ favor the appearance of squeezing as a function of the time. However, independent of the preparation of the initial state, it evolves to an asymptotic steady state which behaves as an ISS, i.e. $\zeta^2_{x'}=-\zeta^2_{y'}$ [dB].
To understand the nature of this asymptotic steady ISS, we have studied the
dependence, as a function of time, of polar angle of the unit vector along the direction of the mean value of the quasi-spin operator $<{\bf S}> $. The corresponding results are  shown in Figure 2. The parameters are the same of those of Figure 1.
In Insets (a) and (b) we have displayed the results obtained when the initial coherent state is prepared with $\left( \theta_0, \phi_0 \right)=(\pi/4,0)$, and with  $(\pi/8,0)$, respectively. The system evolves to an state with $<{\bf S}>$ pointing in the $z$-direction, with $<S_z>=-S$, independent of the choice in the initial coherent state.

In Figure 3, we show
the contribution of the $k-$th state of the basis ${\mathcal A}_k$ to the state $|I(t) \rangle$ of Eq.(\ref{ini00}), as a function of time,
$w(k)=|\langle k| I(t) \rangle|^2$. We have adopted the same parameters of are those of Figures 1 and 2.
In Insets (a) and (b) are displayed the results obtained when the initial coherent state is prepared with $\left( \theta_0, \phi_0 \right)=(\pi/4,0)$ and $(\pi/8,0)$, respectively. From the analysis of Figure 3, it can be concluded that as the state evolves in time, the dominant contributions to the state come from the channels with low values of $k$. This fact is in correspondence with the results of Figure 2.

In Figure 4, we show the dependence, as a function of the relative coupling constant $\eta$, of the squeezing parameters of the steady state ($t>> T_c$, $t=120$ [$\mu$ sec]), in units of [dB]. In Insets (a), (b) and (c) we study systems with $N=5$, $N=45$ and $N=101$ spins, respectively. With solid lines we show the results which we have obtained from the exact diagonalization of the Hamiltonian of Eq.(\ref{hota}), for and initial coherent state with $(\theta_0,~\phi_0)=(\pi/4,~0)$, Eq. (\ref{istate}). The dotted-line is used to show the behaviour of $\zeta^2_{x'} \times \zeta^2_{y'}$ in units of [dB].
The results presented support the idea of the existence of to regions with
different squeezing properties. The initial coherent state evolves into a steady ISS for $\eta < 1 $, and looses the squeezing properties if $\eta > 1 $. In the next sections, we shall present some analytical results to understand this property, and we shall discussed the rest of the curves of the Figure.

Next, we shall study the persistence of an steady ISS as the number of spins is increased. Figure 5 shows the behavior of the squeezing parameters of the steady state, $\zeta^2_{x'}$ and $\zeta^2_{y'}$, as a function of the number of spins of the system, in units of [dB]. The curves have been computing at instant $t=120$ [$\mu$ sec], with  $t>> T_C$. In Insets (a), (b), (c) and (d) we show the results that we have obtained when the relative coupling constant $\eta$ takes the value $\eta=0.25$, $\eta=0.50$, $\eta=0.75$ and $\eta=0.95$, respectively. We have chosen an initial coherent state with $(\theta_0,~\phi_0)=(\pi/4,~0)$. The rest of the parameters are those of Figure 1. We have plotted with circles the value of the product $\zeta^2_{x'}~\zeta^2_{y'}$ in units of [dB]. The line at constant value $0$, is just plotted as a guide. As it can be observed from the Figure, except for systems with small number of spins at large values of $\eta$, the steady state behaves as an ISS. Also, it can be observed that the amount of squeezing achieved in the steady state is increased as the value of the relative coupling constant approaches $\eta \rightarrow 1$.

\begin{figure}[h!]
\includegraphics[width=8cm]{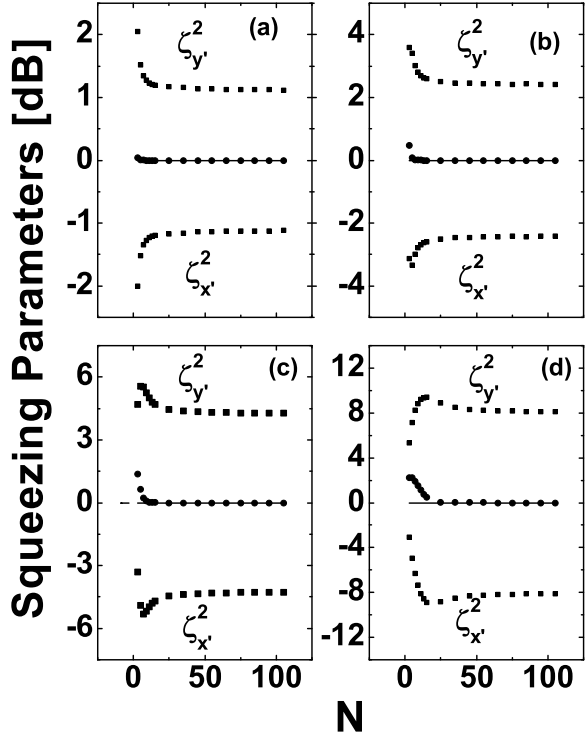}
\caption{Behavior of the Squeezing Parameters of the steady state, $\zeta^2_{x'}$ and $\zeta^2_{y'}$, as a function of the number of spins, $N$ ($t>> T_c$, $t=120$ [$\mu$ sec]), in units of [dB]. In Insets (a), (b), (c) and (d) we show the results obtained when the relative coupling constant $\eta$ is fixed to the value $\eta=0.25$, $\eta=0.50$, $\eta=0.75$ and $\eta=0.95$, respectively. The rest of the parameters are those of Figure 1. We have plotted with circles the value of the product $\zeta^2_{x'}~\zeta^2_{y'}$ in [dB]. The line at constant value $0$, is just to guide the eye.} \label{fig:fig5}
\end{figure}

In what follows we shall present some analytical results in order to understand the behaviour of the steady state of the system as an ISS.

\subsection{Non-hermitian OAT model.}\label{otwist}
Let us first consider the time evolution of the initial state proposed in Eq.(\ref{istate}), under the Hamiltonian
\beqn
H_0 & = & \chi \, S_z^2 + \left( \epsilon-{\rm \bf i} \gamma \right) \left(S_z+ S \right),
\label{hota0}
\eeqn
that is in absence of the LMG interaction. The mean values of the spin components can be calculated straightforwardly, and they read

\beqn
\langle S_z \rangle   & = & - S \frac{ 1- |\widetilde{z}|^2 }  { 1+  |\widetilde{z}|^2}, \nonumber \\
\langle S_z^2 \rangle & = & S^2- \frac{ 2 S (2 S-1)  |\widetilde{z|}^2 }{ (1+  |\widetilde{z}|^2)^2}, \nonumber \\
\langle  \{ S_+,S_-  \} \rangle & = & 2 S +
\frac{4 S(2 S-1) |\widetilde{z}|^2 }{ \left(1+ |\widetilde{z}|^2\right)^2} , \nonumber\\
\langle S_+ \rangle & = & 2 S  \widetilde{z}^* e^{i \epsilon t }
     \frac{ \left(e^{-i t \chi } +   |\widetilde{z}|^2 e^{i t \chi }\right)^{2 S-1}}
    {\left(1+  |\widetilde{z}|^2 \right)^{2 S}},  \nonumber \\
\langle S_+^2\rangle & = & 2 S (2 S-1) \widetilde{z}^{* 2}  e^{i 2 \epsilon t }
\nonumber \\
& & ~~~~~~~~~~~~~
\frac {\left( e^{-2 i  \chi t} +|\widetilde{z}|^2  e^{2 i  \chi t }\right)^{2 (S-1)}}
{\left( 1 +|\widetilde{z}|^2 \right)^{2 S}}, \nonumber \\
\eeqn
being $\widetilde{z}=z(\theta_0,\phi_0) e^{-\gamma t}$.

Clearly, $\widetilde{z}\rightarrow 0$ when $t\rightarrow\infty$. In this limit we find

\beqn
\langle S_z\rangle   & \rightarrow & - S , \nonumber \\
\langle S_x\rangle = Re(\langle S_+\rangle)& \rightarrow & 0  , \nonumber\\
\langle S_x^2\rangle = \frac 12 Re(\langle S_+^2\rangle)+ \frac 14 \langle  \{ S_+,S_- \} \rangle & \rightarrow & \frac S 2,\nonumber \\
\langle S_y\rangle= Im(\langle S_+\rangle)& \rightarrow & 0,  \nonumber \\
\langle S_y^2\rangle= -\frac 12 Re(\langle S_+^2\rangle)+ \frac 14 \langle  \{ S_+,S_- \} \rangle & \rightarrow & \frac S 2. \nonumber \\
\eeqn
Consequently, $\langle{\bf S}\rangle\rightarrow - S \breve{e_z}$, with

\beqn
\Delta^2 S_x  \rightarrow  \frac S 2, ~~~~~
\Delta^2 S_y  \rightarrow  \frac S 2.
\eeqn
This results indicates that, as reported in the previous Section, the initial coherent spin state,$|I (\theta_0,\phi_0) \rangle$,  evolves, asymptotically, to the state with $|I(\pi,0)\rangle=|S,-S \rangle $, independent of the orientation of the state at $t=0$.

\subsection{Non-hermitian LMG model.}\label{su11}

The purpose of this section it is to provide an analytical hamiltonian which accounts for the behaviour of the system in the stationary regime,
when the Lipkin interaction is included.

We shall perform a Holstein-Primakoff
boson mapping \cite{marshalleck,ring,nosbos} of the Hamiltonian of Eq.(\ref{hota}). The generators of the $su(2)$, in terms of the boson creation operator, $\bp$,  and of the boson
annihilation operator, $b$, read

\beqn
S_+ & = & \bp ~ \sqrt{2 S - \bp b} ~ \approx ~\sqrt{2 S} ~ \bp , \nonumber \\
S_- & = & \sqrt{2 S - \bp b} ~b ~ \approx ~\sqrt{2 S} ~  b, \nonumber \\
S_z & = & \bp b-  S.
\label{sps}
\eeqn
The nonlinearity introduced by the square-root term in Eq. (\ref{sps})
ensures that no two excitations can take place at the same
spin. If we consider delocalized spin waves involving a large
number of spins compared to the number of excitations, the
probability that a given spin is excited is inversely proportional
to the number of spins N. Therefore, as long as only a few
delocalized spin excitations are considered, it is reasonable to
neglect the square-root term in Eq. (\ref{sps})\cite{molmer}.

The assumption we have made in Eq. (\ref{sps}) is valid after the system has reached the stationary regime, and is consistent with the results we have presented in Figure 3.
In this approximation, the Hamiltonian of Eq.(\ref{hota}) can be written as

\beqn
H_{B} & = & h_0+ 2 \alpha K_{0}+ 2 S V (K_{+}+ K_{-}), \nnu
\label{hbos}
\eeqn
with
\beqn
K_+ & = & \frac 1 2  {\bp}^2, ~ K_-=K_+^\dagger \nonumber \\
K_0 & = & \frac 1 2 \bp b + \frac 1 4,
\label{opsu11}
\eeqn
and

\beqn
h_0 & =& \chi S^2-\frac 12 \alpha, \nnu
\alpha & = &  \left( \epsilon- 2 S \chi-{\rm \bf i} \gamma \right).
\eeqn

The set of operators $\{K_+,~K_-,~K_0 \}$ spans the algebra of $su(1, 1)$, that is

\beqn
\left[ K_{-}, K_{+}\right] & = & 2 K_{0}, \nonumber \\
\left[ K_{0}, K_{\pm}\right] & = & \pm K_{\pm}. \\
\label{algsu11}
\eeqn

The time evolution operator of the system, $U(t)=e^{- \uni t H_{B}}$, can be easily computed if the exponential were written in a
normally ordered form  \cite{gilmore,romina}. Making use of the faithful matrix representation of the operators $su(1,1)$-algebra, it reads
(see Appendix)

\begin{eqnarray}
U(t) & = & e^{- \uni t H_{B}} \nnu
     & = & e^{- \uni t h_0}e^{b_{+}K_{+}}e^{\ln(b_{0})K_{0}}e^{b_{+}K_{-}},
\label{tevsu11}
\end{eqnarray}

with
\beqn
b_{0}& = & \left( \cos( t \beta) \left( 1 + \frac{ \alpha}{  \beta} {\rm tanh}(\uni t \beta) \right) \right)^{-2} \nnu
b_{+}& = &  {\rm e}^{\uni (\phi_V+\pi)}\frac{  2 S |V|}{ \beta }
 \frac{{\rm tanh}\left (\uni t \beta \right)}{1+\frac{ \alpha}{\beta} {\rm tanh}\left(\uni t \beta \right)} \nnu
\label{defi}
\eeqn
where, $\phi_V=0$ if $V>0$ and  $\phi_V=\pi$ if $V<0$. We have defined the complex parameter $\beta=\sqrt{\alpha^{2}-(2 S V)^{2}}$.

As $|b_+|<1$ (see Appendix), we can introduce the squeezing parameter $\zeta= r {\rm e}^{\uni (\phi+\phi_V+\pi)}$, such that
\beqn
b_+=(\zeta/|\zeta|)\tanh |\zeta|.
\eeqn

In what follows, we shall study the evolution of the state
\beqn
|\psi \rangle= {\cal N} \sum_{n=0}^{2 S}\frac{(\sqrt{2 S})^{n}}{\sqrt{n!}}|n \rangle=D(\sqrt{2 S})|0 \rangle,
\label{inibos}
\eeqn
where, $D(\eta)={\rm e}^{(\eta b^{\dagger}-  \overline{\eta}b)}$ is the displacement operator. The proposed initial state of Eq.(\ref{inibos}) is the limit to dominant order in the number of spins of the coherent state of Eq.(\ref{istate}). This state evolves in time as (see Appendix)

\beqn
U|\psi \rangle & = &\mathcal{N} {\rm e}^{- \uni t h_0}{R_{0}}^{(1/4)} {\rm e}^{S (|R_{0}|+R_{-}-1)}S_q(\zeta)D(\sqrt{2 S R_{0}})|0\rangle, \nnu
\mathcal{N}^{-2}&=& \langle \psi| U^{\dagger} U | \psi \rangle = {\rm e}^{\gamma t } {\rm e}^{2 S (|R_{0}|+ {\rm Re}(R_{-})-1)}\sqrt{|R_{0}|}.
\label{fit}
\eeqn
The parameters $R_0$ and $R_-$ are given by
\beqn
R_{0} & = & \frac{b_{0}}{1-|b_+|^2}, \nnu
R_{-} & = & \overline{b_+}~R_0 -b_+,
\eeqn
and  $S_{q}(\zeta)$ stands for the squeezing operator, ${S_{q}(\zeta)= {\rm e}^{\overline{\zeta} K_-\zeta K_{+}}}$.

We are, now, in condition to compute the uncertainty relations of the operators

\beqn
x & = & \frac {1}{\sqrt{2}} \left( b^{\dagger}+b \right), \nnu
p & = & \uni \frac {1}{\sqrt{2}} \left( b^{\dagger}-b \right), \nnu
\label{xp}
\eeqn
on the state of Eq.(\ref{fit}). After some cumbersome  algebra (see Appendix) it can be probed that

\beqn
\Delta^{2}x & = & \frac{1}{2} \left(- \cos ( \phi + \phi_V ) \frac{2 \rho}{1-\rho^2}+\frac{1+\rho^2}{1-\rho^2}\right), \nnu
\Delta^{2}p & = & \frac{1}{2} \left(+ \cos ( \phi + \phi_V ) \frac{2 \rho}{1-\rho^2}+ \frac{1+\rho^2}{1-\rho^2} \right),
\label{urs}
\eeqn
with $\rho=|b_+|=\tanh |\zeta|$.
Consequently we can defined the associated squeezing parameters $Q(x,p)$ and $Q(p,x)$ as
\beqn
Q(x,p) = 2 \Delta^2 x, ~~~
Q(p,x) = 2 \Delta^2 p.
\label{sqbos}
\eeqn
The system is squeezed in $x$ ($p$) when $Q(x,p)<1$ ($Q(p,x)<1$).

Our objective is to study the behaviour of the system after a long interval of time ($t \rightarrow \infty$).

Due to decoherence, it is straightforward to show that

\beqn
\lim_{ t \rightarrow \infty} b_+ =
\frac{ {\rm e}^{\uni (\phi_V)} \eta} {\sqrt{ (1- \uni \sigma \Gamma)^2-\eta^2}-(\sigma + \uni \Gamma )}={\rm e}^{\uni (\phi_V+\phi)} \rho_L. \nnu
\label{bplusasymp}
\eeqn
In the previous expression, $\sigma$ stands for the sign
function of ${(\epsilon -2 S \chi)}$,  and

\beqn
\phi & = & -\arctan \left (\frac{\beta_{-}-\Gamma}{\beta_{+}-\sigma}\right)\nnu
\rho_L & = & \frac{\eta}{\sqrt{(\beta_{+}-\sigma)^{2}+(\beta_{-}-\Gamma)^{2}}}.\nnu
\label{phi}
\eeqn
being

\begin{eqnarray}
\beta_{\pm}^2 & = & \frac 1 2 \left(  \sqrt{(1-\eta^{2}-\Gamma^{2})^{2}+ 4 \Gamma^{2}} \pm (1-\eta^{2}-\Gamma^{2})  \right), \nnu
\end{eqnarray}
To leading order in $\Gamma$, the phase factor $\phi$ can be written as

\beqn
\phi \approx \left \{
\begin{array}{ll}
\arctan \left ( \frac {\Gamma}{\sqrt{1-\eta^2}-\sigma} \right ),& 0<\Xi< 1\nnu
\arctan \left (\sigma (\sqrt{\eta^2-1}-\Gamma)\right),& \Xi>1. \nnu
\end{array}
\right.
\label{faseaprox}
\eeqn
We can identify two regions, in the space of coupling constants $\eta$ and $\Gamma$, with different squeezing properties for the steady state of the system.
Region I corresponds to values of $\eta$ and $\Gamma$ that satisfy the condition
$\Xi^2<1 $,  and Region II for values of $\eta$ and $\Gamma$ that satisfy $\Xi^2>1$.

In Region I, for small values of $\Gamma$, the phase $\phi$ of $b_+$ becomes approximately null, $\phi~<<~1$, so that $ b_+  \cong  {\rm e}^{\uni (\phi_V)} \rho_{L} $. Then,
the uncertainty relations of the operators $x$ and $p$, of Eq.(\ref{urs}), for $\phi_V=0$, take the form

\beqn
Q(x,p) & \rightarrow & \frac {1+\rho_L}{1-\rho_L}={\rm e}^{ 2|\zeta|}, \nnu
Q(p,x) & \rightarrow & \frac {1-\rho_L}{1+\rho_L}={\rm e}^{-2|\zeta|} ,
\label{urs0}
\eeqn
and
 \beqn
Q(x,p) Q(p,x) & \rightarrow & 1.
\eeqn
Thus, the steady state of the system behaves as an ISS.
Similar expressions hold for $\phi_V=\pi$, but with the exchange of the roles of $x$ and $p$.

In Region II, the behaviour of the system is completely different.
The phase $\phi$ of $b_+$ is no longer null,  $\phi \ne 0$, moreover for values of $\eta$ sufficiently large $\phi \rightarrow \pm \pi/2$, depending on $\sigma$.
In this case the uncertainty relations of the operators $x$ and $p$, of Eq.(\ref{urs}) take the form

\beqn
Q(x,p) & \rightarrow &  \frac {1+\rho_L^{2}}{1-\rho_L^{2}}, \nnu
Q(p,x) & \rightarrow &  \frac {1+\rho_L^{2}}{1-\rho_L^{2}}.
\label{urs0}
\eeqn
Thus, in Region II, the asymptotic steady state is not a squeezed state.

Let us compared these analytical results with the ones discussed in Section \ref{numbers}.

In view of Eq.(\ref{sps}) and of Eq. (\ref{xp}), to leading order in the number of spins

\beqn
\frac {S_x} {\sqrt{S}} & = & \frac {1}{2\sqrt{S}} (S_+ + S_-) \rightarrow x, \nnu
\frac {S_y} {\sqrt{S}} & = &-\frac {\uni}{2\sqrt{S}}    (S_+ - S_-) \rightarrow - p. \nnu
\eeqn
So that when, under the action of the Hamiltonian of Eq. (1), the initial state of Eq.(\ref{istate}) evolves  to a steady state which points in the
$z$-direction, the squeezing parameters $\{ \zeta^2_{x'}, \zeta^2_{y'} \}$ should  give the same information as $\{ Q(x,p), Q(p,x)\}$.

\begin{figure}[h!]
\includegraphics[width=8cm]{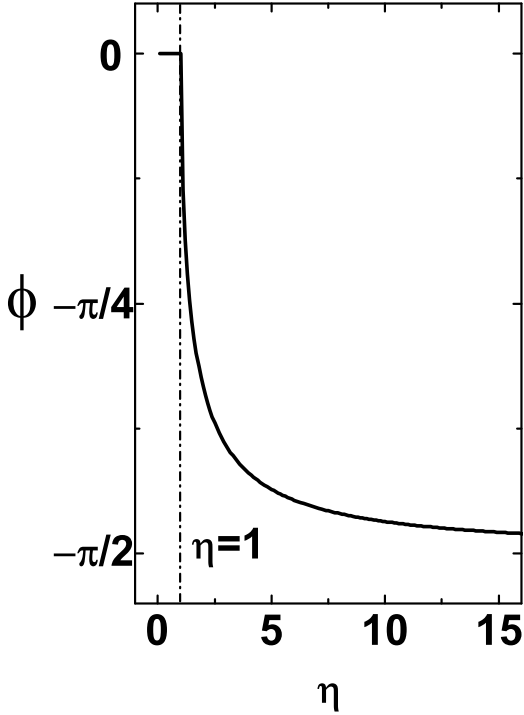}
\caption{ Behaviour of the phase $\phi$ of Eq. (\ref{phi}) as a function of the scaled coupling constant $\eta$. The values of the different parameters are those of Figure 6.} \label{fig:fig6}
\end{figure}


 This can be seen from  Figure 4, where we present, by using dashed-lines, the results obtained for $Q(x,p)$ and $Q(p,x)$ of Eq.(\ref{sqbos}), for the coherent state of Eq. (\ref{inibos}) with $N$ particles in mean value. With dashed-dotted-line we present the results for the product of the squeezing parameters in units of [dB].  Clearly, for systems with more than 9 spins, the initial coherent state evolves into a steady ISS for $\Xi^2< 1 $, and looses the squeezing properties if $\Xi^2 > 1$.

We complete our analytical results by analysing the behavior of the phase $\phi$ of Eq. (\ref{phi}). The results are presented in Figure 6, for the same parameters of Figure 5. The numerical results are in agreement with the analytical estimations  of Subsection \ref{su11}. That is, in Region I the phase $\phi$ is null, $\phi=0$ and consequently the steady state is an ISS, while in Region II $\phi \rightarrow -\pi/2$ for increasing values of the coupling constant $\eta$, and the steady state is no longer an ISS.

From the presented results it can be inferred that dissipative mechanisms can be used to improve the achievement of squeezing in different spin system \cite{disi-0,disi-1}.

\subsection{Application to phonon-induced spin-spin interactions in diamond nanostructures.}\label{application}

Let us consider the spin-spin interaction, among NV centers in diamond, mediated through the coupling of the spins to a magnetic nano-resonator \cite{phonon1,ma-1,ma-2}.

An NV center has a ground state with spin $1$ and a
zero-field splitting D = 2.88 GHz between the $|1,0>$ and
$|1,\pm 1>$ states \cite{nv-1}. If an external magnetic field, ${\bf B}_0$
along the crystalline axis of the NV center, is applied an additional
Zeeman splitting between $|1, \pm 1>$ sub-levels occurs. Then, it is possible to isolate the subsystem  $|1,0 \rangle$ and $|1,-1 \rangle$\cite{marco,disi-1,ma-1,ma-2}.

The mechanical resonator is described by the Hamiltonian $H_r= ~ \omega_r~b^\dagger b$, with $\omega_r$ as the frequency of the fundamental mode vibration mode of the resonator, and $b$ ($b^\dagger$) as the corresponding annihilation (creation) operator. We shall chose $\omega_r$ almost in resonance with the splitting of the states $|1,0\rangle$ and $|1,-1 \rangle$, so that the NV center can be modeled by a two-level system. The motion of the magnetic mechanical resonator produces a magnetic gradient field on the NV centers, so that within this two-level subspace the Hamiltonian of the system can be modeled as

\begin{eqnarray}
H_{NV}= ~ \omega_r~b^\dagger b +~ \delta ~\sigma_z &+&~ g_1 ~ (\sigma_+ b^\dagger+ b \sigma_-)+ \nonumber \\
     & & ~ g_2 ~ (\sigma_+ b+ b^\dagger \sigma_-),
\label{hnv}
\end{eqnarray}
where $\delta=D-\gamma_e B_0$ is the energy gap between the ground state $|1,0 \rangle$ and the state $|1,-1 \rangle$, being $\gamma_e$ the gyromagnetic ratio of an electron.
We have assumed an asymmetric interaction between the NV centers and the single mode mechanical resonator, which is model by the effective coupling constants parameter $g_1$ and $g_2$.
The operators ${\sigma_x,\sigma_y, \sigma_z}$ are collective spin operators for the ensemble of NV centers in diamond, $\sigma_\alpha=\sum_i~\sigma_{\alpha~i}$, which satisfy the usual angular momentum commutation relations.
We shall consider that the intensity of the external magnetic field is fixed in order to have a detuning $\delta \approx 0$.

A unitary transformation of the form
$$U={\rm  e}^{-(g_1/\omega_r) (\sigma_+ b^\dagger-b \sigma_-)~-(g_2/\omega_r) (\sigma_+ b-b^\dagger \sigma_-)}$$ can be applied to the Hamiltonian  of Eq. (\ref{hnv}), $H_{eff}=U H U^{-1}$.
To leading order in $g_1/\omega_r$ and $g_1/\omega_r$, together with the assumption that $\delta \approx 0$, the effective Hamiltonian takes the form

\beqn
H_{eff} \approx ~H_0 + \omega_r~b^\dagger b +2\frac{ g_1^2-g_2^2}{\omega_r}(1+2 b^\dagger b)\sigma_z + \nonumber \\
 2\frac{ g_1^2+g_2^2}{\omega_r}~\sigma_z^2-4\frac{ g_1 g_2}{\omega_r}~ \left({\sigma_x}^2-{\sigma_y}^2 \right), \nonumber
\label{heff}
\eeqn
with $H_0=-2\frac{g_1^2+g_2^2}{\omega_r} S (S+1)$.
We shall account for dissipation by introducing the mean-life of the NV centers through the additional term

\beqn
H_\gamma= -{\rm \bf i} \gamma \left(\sigma_z+ S \right).
\eeqn
The characteristic time of coherence of this system
is of the order of $T_C=100$ [$\mu$ sec ]\cite{zhu,nv-ct1,nv-ct2,nv-ct3}, which is consisting with a value for the line-width of the states
of $\gamma=2 \times 10^{-5}$ [GHz].
Thus, the Hamiltonian of the NV ensemble reads

\beqn
H_{NVE-ph}=H_{eff}+H_\gamma.
\eeqn

In order to generate a steady ISS, we initialize the ensemble of NV centers in a coherent state (CSS) $| CSS \rangle $ along the direction
$\breve{n}_0=(\sin (\theta_0)\cos(\phi_0),\sin (\theta_0)\sin(\phi_0),\cos (\theta_0))$ of the collective Bloch sphere. As it is well known, the CSS satisfies the condition ${\bf \sigma}.\breve{n}_0 | CSS \rangle= -S | CSS \rangle$, and it has equal transverse variances, $S/2$. This state can be prepared by using optical pumping and microwave spin manipulation applied to the ensemble \cite{phonon1,natphys4}.

The Hamiltonian of Eq. (\ref{heff}) includes a term which couples the phonon number $\hat{n}=a^\dagger a$ to $\sigma_z$. We shall consider an initial phonon with $<\hat{n}>=n_{ph}$, which we shall model as a coherent sate of the form
$$ |n_{ph} \rangle= {\rm e}^{-|z_{ph}|^2/2} \sum_{n=0}^{\infty} \frac{z_{ph}^n}{\sqrt{n!}} |n \rangle,$$
where $|n \rangle$ represents the state with $n$ phonons, and $|z_{ph}|^2=n_{ph}$. An initial state of the form
 $|I\rangle= |n_{ph} \rangle |CSS \rangle $, will evolve as

\begin{eqnarray}
|I(t)\rangle = {\rm e}^{-|z_{ph}|^2/2} \sum_n ~ \frac{z_{ph}^n}{\sqrt{n!}}|n\rangle ~|I_{NVE} (t,n) \rangle,
\end{eqnarray}
with
\begin{eqnarray}
|I_{NVE} (t,n) \rangle = {\rm e}^{{\bf i} H_{NVE}(n)t} |CSS \rangle,
\end{eqnarray}
and
\begin{eqnarray}
H_{NVE} (n) & = & \epsilon \sigma_z + \nonumber \\
 & & \chi~\sigma_z^2 + V~ \left({\sigma_x}^2-{\sigma_y}^2 \right)+H_{\gamma},  \nonumber\\
 \epsilon & = & 2\frac{ g_1^2-g_2^2}{\omega_r}(1+2 n )\sigma_z,\nonumber \\
 \chi& = & 2\frac{ g_1^2+g_2^2}{\omega_r}~\sigma_z^2, \nonumber \\
 V & = & -4\frac{ g_1 g_2}{\omega_r}.
\label{heffnv}
\end{eqnarray}
Following the formalism presented in \ref{tevol}, the mean value of physical operator associated to the NV centers, $\hat{o}_{NV}$, will be computed as

\begin{eqnarray}
\langle \hat{o}_{NV} (t)\rangle
&  = &{\rm e}^{-|z_{ph}|^2} \sum_{n=0}^{\infty} \frac{|z_{ph}|^{2 n}}{n!} ~\langle I_{NVE} (t,n)  |\hat{o}_{NV}|I_{NVE} (t,n) \rangle_{\mathcal S},\nonumber \\
\label{teph}
\end{eqnarray}
where the $\mathcal{S}$ is the corresponding metric operator \cite{arxiv}.

In the previous section we have conclude that, for large number of NV centers, the values of $\Xi=\eta^2+\Gamma^2$ (Eq.(\ref{xi})) can be used to characterize the appearance of a steady ISS, that is for If $\Xi^2<1$, the initial state evolves into a steady ISS. In terms of $g_1,~g_2,~w_r,~\gamma$ and of the number of NV centers, $N=2 S$ and of the number of phonons, $n$, the quantity $\Xi^2$ reads

\begin{eqnarray}
\Xi^2& =&\frac{2 \frac{g_1}{g_2}+ \frac{\gamma}{4 S~(g_2^2/\omega_r)}}
{
\left|  \left( \frac{g_1} {g_2}  \right )^2  \left(1-\frac{1+2 n}{2 S}  \right) + \left(1+\frac{1+2 n}{2 S} \right)\right|}.
                 \nonumber \\
\label{nvph}
\end{eqnarray}
The quantity $\Xi^2$, if $\gamma/(4 S)$ is small, depends on the relative coupling constant $g_1/g_2$ and on the ratio of phonon numbers to the number of spins, $(1+ 2n)/(2 S)$.

In Figure 6, we present a contour plot of $\Xi^2$ as a function of the ratios  $g_1/g_2$ and  $(1+ 2n)/(2S)$. We have considered a system of $N=1001$ NV-centers. We have taken values of $\omega_r=1$ [MHz], $g_2=0.5$ [MHz] and $\gamma=2 \times 10^{-2}$ [MHz] \cite{phonon1,ma-1,ma-2}. From the Figure it can be seen that $\Xi^2 <1$ for values of $g_1/g_2<1$, or for $(1+ 2n)/(2 S) \lesssim 0.5$ if $g_1/g_2$. Similar results are obtained for systems with different values of the number of the NV centers, $N=2 S$, and of the number of phonons, provided that $(1+ 2n)/(2S)$ varies among the same values.

\begin{figure}[h!]
\includegraphics[width=8cm]{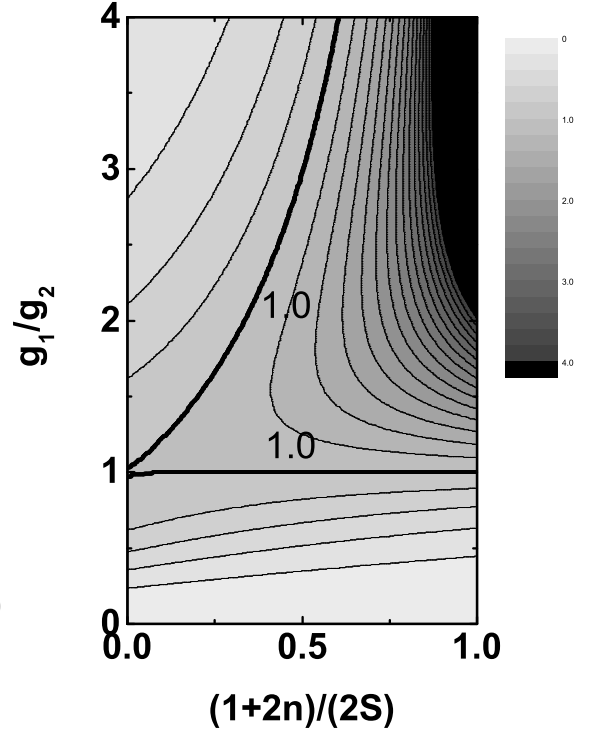}
\caption{Contour plot of the quantity $\Xi^2$, as a function of the ratios  $g_1/g_2$ and  $(1+ 2n)/(2S)$. We have considered a system of $N=2 S=1001$ NV-centers. We have taken values of $\omega_r=1$ [MHz], $g_2=0.5$ [MHz] and $\gamma=2 \times 10^{-2}$ [MHz]. } \label{fig:fig7}
\end{figure}

In Figure 7, we show the results obtained for the squeezing parameter of the steady state, as a function of the ratio $g_1/g_2$. We have computed the mean values of the physical operators following Eq. (\ref{teph}). We have chosen an initial coherent state for the NV centers, with $\theta_0=\pi/4$ and $\phi_0=0$.
We have considered a system with $N= 2 S=1001$ NV color centers in diamond. The values of $g_2$, $\omega_r$ and $\gamma$ are those of Figure 8. We have evaluated the Squeezing parameter at $t=300$ [$\mu$ s] $>> T_C$. In Insets (a), (b) and (c) we show the results obtained when the mean value of phonons in the initial state, Eq.(\ref{inibos}), is $n_{ph}=~6,~100$ and $250$, respectively. When the mean value of phonons is increased, the contribution from states with large number of $n$ becomes important, so that, at fix number of NV-centers, the parameter $\Xi^2$ can be $>1$ depending on the ratio $g_1/g_2$.
We have verified that the values of the squeezing parameter in the steady state are independent of the initial state adopted \cite{ma-2}.

\begin{figure}[h!]
\includegraphics[width=8cm]{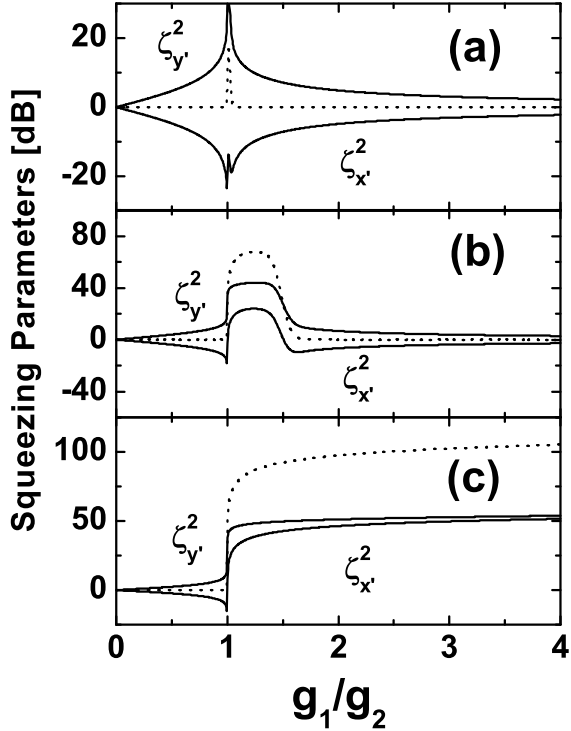}
\caption{Values of the Squeezing Parameters in the steady state, as a function of the relative constant $g_1/g_2$, in units of [dB]. We have considered a system with $N= 2 S=10001$ NV color centers in diamond. The values of $g_2$, $\omega_r$ and $\gamma$ are those of Figure 7. We have evaluated the Squeezing parameter at $t=300$ [$\mu$ s] $>> T_C$. In Insets (a), (b) and (c) we show the results obtained when the mean value of phonons is $n_{ph}=~6,~100$ and $250$, respectively.} \label{fig:fig8}
\end{figure}

\section{Conclusions}\label{conclusions}

In this work we have studied the behavior of a system of spins interacting through a non-hermitian one-axis
twisting Hamiltonian plus a Lipkin-type interaction.
We have analysed the time evolution of a coherent initial spin state.
We have shown, by performing the exact numerical diagonalization of the Hamiltonian, that under the action of the one-axes twisting dissipative hamiltonian, the initial state evolves into steady coherent state pointing in the z-direction.
This fact have been proved analytically in Section \ref{otwist}.
In addition, in Section \ref{numbers} we have shown that, by performing an exact diagonalization of the interaction of Hamiltonian (\ref{hota}), when the Lipkin interaction is turned on, a coherent initial state evolves into steady
Intelligent Spin State for a definite range of values of the relative coupling constant $\eta$.
To get a deeper understanding of the results we have obtained in the staionary regime, we have performed a boson mapping of the $su(2)$ Hamiltonian of Eq.(\ref{hota}). To leading order in the number of spins, the
Hamiltonian was written in terms of the operators of the $su(1,1)$ algebra, and the time evolution of the system was
obtained analytically.
In the asymptotic limit, that is after long intervals of time compared to the characteristic coherence time of the system,
the numerical results that we have presented support the idea that the behaviour of the steady state govern by the $su(2)$-Hamiltonian
of Eq. (\ref{hota}) can be understood in terms of the behavior of the steady state govern by the $su(1,1)$-Hamiltonian of Eq. (\ref{hbos}).
Both from analytical and numerical results, it is observed that two well defined regions can be identified,
depending on the relative value of the coupling constants ($\eta,~ \Gamma$), with different behaviour of the asymptotic steady state.
For systems with more than $N \approx 9$ spins,
the initial state evolves in a steady Intelligent Spin State when $\eta<1$, Eq. (\ref{xi}), otherwise the  asymptotic state does not behave as a squeezed state.
The previous reported results indicate  that the generation of a steady Intelligent Spin State, for a certain range of values of ($\eta$, $\Gamma$), is consequence of the dissipative character of the interaction. Similar results have been recently advanced in \cite{disi-1}. As a potential physical application, we have investigated the possibility of searching for an steady Intelligent Spin State in diamond nano-structures. We have presented an effective spin-spin interaction among NV color centers in diamond, mediated through the interaction of the NV centers with a magnetic nano-resonator. We have investigated the regimen of coupling constatnts, so that under the action of this effective interaction a coherent initial state evolves in time into a steady Intelligent Spin State.

\begin{acknowledgments}
This work was partially supported by the National Research Council
of Argentine (PIP 282, CONICET) and by the Agencia Nacional de
Promocion Cientifica (PICT 001103, ANPCYT) of Argentina.
\end{acknowledgments}

\section*{Appendix}\label{appendix}

Let us consider the Lie algebra $su(1, 1)$ \cite{romina}, which is spanned by the operators $\{ K_1,~K_2,~K_3 \}$.  They satisfy
the well known commutation relations
$$[K_1,K_2] = - \uni K_3, \ \ [K_2,K_3] = \uni K_1, \ \  [K_3,K_1] = \uni K_2.$$
The complex linear combinations of these operators span the algebra $su^{c}(1, 1)$, which is isomorphic to $sl(2,C)$.

The time evolution operator, $U(t)=e^{- \uni t H_{B}}$, is an exponential form of the elements of the $su^c(1,1)$ Lie algebra
( Eqs. (\ref{opsu11}) and (\ref{algsu11})). Thus, $U(t)$ belongs to the $SU(1,1)$ Lie group.  Consequently, $U(t)$ can be represented by a matrix $G$. The matrix $G$
is  parameterized by two complex numbers  $w_{1}$ and $w_{2}$ as
$$G=\left(
      \begin{array}{cc}
        w_{1} & w_{2} \\
        \overline{w}_{2} & \overline{w}_{1} \\
      \end{array}
    \right),
$$
moreover, the parameters $w_1$ and $w_2$ fulfill the condition $|w_{1}|^{2}-|w_{2}|^{2}=1$.

Let us determine $w_1$ and $w_2$. In doing so, we shall write $U(t)$ in normal order as

\begin{eqnarray*}
U(t) & = & e^{- \uni t h_0} e^{- \uni t (2 \alpha K_{0}+ 2 S V (K_{+}+ K_{-}))},\\
     & = & e^{- \uni t h_0} e^{b_{+}K_{+}}e^{\ln(b_{0})K_{0}}e^{b_{+}K_{-}},
\label{rightut}
\end{eqnarray*}
with $ K_{\pm}  =  K_{1} \pm \uni K_{2}$ and $K_0=K_3$.

Following the prescriptions of \cite{gilmore}, it is possible to carry out all calculations, in either the algebra or
the group, by using the faithful matrix representation of the operator algebra. It reads
\begin{eqnarray*}
K_{+} & = & \left(
          \begin{array}{cc}
            0 & 1 \\
            0 & 0 \\
          \end{array}
            \right), \nnu
K_{-} & = & \left(
          \begin{array}{cc}
            0 & 0 \\
            -1 & 0 \\
          \end{array}
        \right), \nnu
K_{0} &= & \frac{1}{2}\left(
          \begin{array}{cc}
            1 & 0 \\
            0 & -1 \\
          \end{array}
        \right).
\end{eqnarray*}
Writting (\ref{rightut}) in terms of the faithful matrix representation, we obtain

\begin{eqnarray*}
\left(
     \begin{array}{ll}
       c+ \frac{a_{0}s}{2 d} & \frac{a s }{d} \\
       - \frac{a s }{d}  & c- \frac{a_{0}s}{2 d} \\
     \end{array}
   \right)
=
\left(
     \begin{array}{ll}
       \sqrt{b_0}-\frac {b_+ b_-}{\sqrt{b0}} & \frac {b_+}{\sqrt{b0}} \\
      -\frac { b_-}{\sqrt{b0}}& \frac {1}{\sqrt{b0}} \\
     \end{array}
   \right),
\end{eqnarray*}
where

\begin{eqnarray*}
c & = & \cosh( \uni t \beta ), \\
s & = & \sinh( \uni t \beta), \\
a_{0} & =& -2 \uni t \alpha, \\
a & =& - \uni t \gamma.
\end{eqnarray*}
Then, it results
\begin{eqnarray*}
b_{0}& = & \left( \cos( t \beta) \left( 1 + \frac{ \alpha}{  \beta} {\rm tanh}(\uni t \beta) \right) \right)^{-2}, \nnu
b_{+}& = &  {\rm e}^{\uni (\phi_V+\pi)}\frac{  2 S |V|}{ \beta }
 \frac{{\rm tanh}\left (\uni t \beta \right)}{1+\frac{ \alpha}{\beta} {\rm tanh}\left(\uni t \beta \right)}, \nnu
b_{-}& = & b_+,
\label{defi}
\end{eqnarray*}
where, $\phi_V=0$ if $V>0$ and  $\phi_V=\pi$ if $V<0$. We have defined $\beta = \sqrt{\alpha^{2}-(2 S V)^{2}}$.
Clearly,  we can identify

\begin{eqnarray*}
\begin{array}{l l}
w_{1} = \sqrt{b_{0}}-\frac{b_{+}^{2}}{\sqrt{b_{0}}} &, \overline{w}_{1}=\frac{1}{\sqrt{b_{0}}},\\
w_{2} = \frac{b_{+}}{\sqrt{b_{0}}}                  &, \overline{w}_{2}=-\frac{b_{+}}{\sqrt{b_{0}}}\\
\end{array}
\end{eqnarray*}
As $|w_{1}|^{2}-|w_{2}|^{2}=1$, there exist  $\zeta \in {\mathcal C}$ and $\{\theta_{1},~ \theta_{2}\}~ \in {\mathcal R}$ so that
$$w_{1}=\cosh |\zeta| {\rm e}^{i \theta_{1}} \ \ \ \ w_{2}=\sinh |\zeta| {\rm e}^{i \theta_{2}}.$$ Consequently

$$ \frac{w_{2}}{\overline{w}_{1}} =b_{+}= {\rm e}^{\uni (\theta_2-\theta_1)}\tanh|\zeta|,$$
verifying that $|b_{+}|<1.$

It is convenient to introduce the operator of squeezing $S_q(\zeta)=e^{\overline{\zeta}K_{-}-\zeta K_{+}}$,
with $\zeta= r {\rm e}^{\uni \tau}$ and $\tau=\phi+\phi_V+\pi$. In terms of the complex parameter $\zeta$, $b_+$ is written as $b_+=\frac {\zeta} {|\zeta|}\tanh |\zeta|$.
It is straightforward to show, by using the faithful matrix representation, that

$$
S_q(\zeta)=e^{\overline{\zeta}K_{-}-\zeta K_{+}}=e^{b_{+}K_{+}}e^{\ln(1-\tanh^2|\zeta|)K_{0}}e^{-\overline{b_{+}} K_{-}},
$$
and

\begin{eqnarray*}
U(t)& = &
e^{- i t h_0}e^{b_{+}K_{+}}e^{\ln(1-\tanh^2|\zeta|)K_{0}}e^{-\overline{b_{+}}K_{-}} \nnu
& & ~~~~~~~~~e^{\overline{b_{+}}K_{-}}e^{-\ln(1-\tanh^2|\zeta|)K_{0}}e^{\ln(b_{0})K_{0}}e^{b_{+}K_{-}}
\nnu
&=&{\rm e}^{- \uni t h_0}S_q(\zeta){\rm e}^{\ln(R_{0})K_{0}}{\rm e}^{R_{-}K_{-}},\nnu
\end{eqnarray*}
where we have defined $R_{0}  = \frac{b_{0}}{1-\tanh ^2|\zeta|} $, and \\
 $R_{-}  =  \left( \frac{\overline{\zeta}}{|\zeta|} \frac{b_{0}}{1-\tanh ^2|\zeta|} -\frac{\zeta}{|\zeta|} \right) \tanh |\zeta|$.

We shall now consider the time evolution of the coherent state of Eq.(\ref{inibos})

\begin{eqnarray*}
|\psi \rangle= e^{-S}\sum_{n=0}^{\infty}\frac{(\sqrt{2S})^{n}}{\sqrt{n!}}|n \rangle=D(\sqrt{2S})|0 \rangle,
\end{eqnarray*}
with $D(\sqrt{2S})=\exp(\sqrt{2S} a^{\dagger}-  \sqrt{2S}a)$. It is easy to proof that $K_{-}|\psi \rangle= S|\psi \rangle$. Then

\begin{eqnarray*}
U|\psi \rangle & = & \mathcal{N} {\rm e}^{- \uni t h_0}S_q(\zeta){\rm e}^{\ln(R_{0})K_{0}}{\rm e}^{R_{-}K_{-}}D(\sqrt{2S})|0 \rangle, \nnu
& = & \mathcal{N} {\rm e}^{- \uni t h_0}S_q(\zeta){\rm e}^{\ln(R_{0})K_{0}}{\rm e}^{S R_{-}}D(\sqrt{2S})|0 \rangle, \nnu
& = & \mathcal{N} {\rm e}^{- \uni t h_0}R_0^{1/4}{\rm e}^{S R_{-}-S+ |R_0|} S_q(\zeta)D(\sqrt{2S R_0})|0 \rangle, \nnu
\end{eqnarray*}
and the normalization factor results

\begin{eqnarray*}
\mathcal{N}^{-2}&=& e^{\gamma t} {\rm e}^{2 S (|R_{0}|+{\rm Re}(R_{-})-1)}\sqrt{|R_{0}|} \times \nnu
& & ~~~~~~~~ \langle 0|D^{\dagger}(\sqrt{2 S R_{0}})S^{\dagger}(\zeta)S_q(\zeta)D(\sqrt{2 S R_{0}})|0\rangle \nnu
&=&e^{\gamma t }e^{2 S (|R_{0}|+Re(R_{-})-1)}\sqrt{|R_{0}|}.
\end{eqnarray*}

Let us evaluate the fluctuation of the operators $x$ and $p$. In doing so, we shall make use of well known relations for the squeezing operator $S(\zeta)$:

\begin{eqnarray*}
S_q^{\dagger}(\zeta) x S_q(\zeta)& = & x~(\cosh r-\cos \tau \sinh r)-p~ \sin \tau \sinh r,\nnu
S_q^{\dagger}(\zeta) p S_q(\zeta)& = & p~(\cosh r+\cos \tau \sinh r)+x~ \sin \tau \sinh r,\nnu
S_q^{\dagger}(\zeta) x^2 S_q(\zeta) & = & x^2~(\cosh r- \cos \tau \sinh r )^{2} + \nnu
                                    &   & p^2~ \sin^2 \tau \sinh^2 r- \nnu
                                    &   & \{x, p\}\sin \tau \sinh r (\cosh r - \cos \tau \sinh r ),\nnu
S_q^{\dagger}(\zeta) x^2 S_q(\zeta) & = & x^2~\sin^2 \tau \sinh^2 r + \nnu
                                    &   & p^2~ (\cosh r+ \cos \tau \sinh r )+ \nnu
                                    &   & \{x, p\}\sin \tau \sinh r (\cosh r + \cos \tau \sinh r ),\nnu
\end{eqnarray*}
and of
\begin{eqnarray*}
\langle 0 |D^{\dagger}(\sqrt{2 S R_{0} })x D(\sqrt{2 s R_{0} })|0 \rangle &=& 2 \sqrt{S} ~ {\rm Re}\sqrt{R_{0}},\nnu
\langle 0 |D^{\dagger}(\sqrt{2 S R_{0} })p D(\sqrt{2 s R_{0} })|0 \rangle &=& 2 \sqrt{S} ~ {\rm Im}\sqrt{R_{0}},\nnu
\langle 0 |D^{\dagger}(\sqrt{2 S R_{0} })x^2 D(\sqrt{2 s R_{0} })|0 \rangle &=& \frac{1}{2}+ 4 S~ ({\rm Re}\sqrt{R_{0}})^{2}\nnu
\langle 0 |D^{\dagger}(\sqrt{2 S R_{0} })p^2 D(\sqrt{2 s R_{0} })|0 \rangle &=& \frac{1}{2}+ 4 S~ ({\rm Im}\sqrt{R_{0}})^{2}.\nnu
\end{eqnarray*}
We can proceed to calculate

\begin{eqnarray*}
\Delta^{2}p & = & \langle \psi | U^{\dagger} p^{2}U | \psi \rangle -  \langle \psi | U^{\dagger} p U |\psi \rangle ^{2}, \nnu
&=& \frac{1}{2} \left(\cos (\tau ) \sinh (2 r)+\sinh ^2(r)+\cosh ^2(r)\right),
\end{eqnarray*}
and
\begin{eqnarray*}
\Delta^{2}x & = & \langle \psi | U^{\dagger} x^{2}U | \psi \rangle -  \langle \psi | U^{\dagger} x U |\psi \rangle ^{2}, \nnu
&=& \frac{1}{2} \left(-\cos(\tau ) \sinh (2 r)+\sinh ^2(r)+\cosh ^2(r)\right).
\end{eqnarray*}

\section*{}

\end{document}